\documentclass[letterpaper, 10 pt, conference]{ieeeconf}
\usepackage{graphicx} % Required for inserting images
\usepackage{float}
\usepackage{subcaption}
\usepackage{amssymb}
\usepackage{tikz}
\usepackage{mathtools}
\usepackage{caption}
\usepackage{scalefnt}
\usepackage{mdframed}

\usepackage[sorting=none]{biblatex}
\title{%
Offline Digital Euro: a Minimum Viable CBDC using Groth-Sahai proofs \\
\large --- Master's Thesis ---
}

\author{Leon Kempen \\ 
    \small Delft University of Technology \\
    \small Delft, The Netherlands \\
    \small L.M.Kempen@student.tudelft.nl \\
   \and
   Johan Pouwelse (thesis supervisor) \\ 
   \small Delft University of Technology \\
   \small Delft, The Netherlands \\
   \small J.A.Pouwelse@tudelft.nl
}

\date{July 2024}
\begin{document}
\maketitle
\begin{abstract}
    Current digital payment solutions are fragile and offer less privacy than traditional cash.
    Their critical dependency on an online service used to perform and validate transactions makes them void if this service is unreachable. 
    Moreover, no transaction can be executed during server malfunctions or power outages. 
    Due to climate change, the likelihood of extreme weather increases. As extreme weather is a major cause of power outages, the frequency of power outages is expected to increase.
    The lack of privacy is an inherent result of their account-based design or the use of a public ledger. 
    The critical dependency and lack of privacy can be resolved with a Central Bank Digital Currency that can be used offline. 
    This thesis proposes a design and a first implementation for an offline-first digital euro. 
    The protocol offers complete privacy during transactions using zero-knowledge proofs.
    Furthermore, transactions can be executed offline without third parties and retroactive double-spending detection is facilitated. 
    To protect the users' privacy, but also guard against money laundering, we have added the following privacy-guarding mechanism. 
    The bank and trusted third parties for law enforcement must collaborate to decrypt transactions, revealing the digital pseudonym used in the transaction. 
    Importantly, the transaction can be decrypted without decrypting prior transactions attached to the digital euro. 
    The protocol has a working initial implementation showcasing its usability and demonstrating functionality.
\end{abstract}
\section{Introduction}
The current infrastructure for digital payment systems is fragile because of their design to support instant payments or to obtain global consensus. However this infrastructure has two main concerns: it is required to be online and allows third parties, other than the payer and payee, to observe balance transfers. A lesser-known digital payment system, electronic cash (e-cash), mitigates these problems. Moreover, e-cash also provides instant payments without requiring a global consensus. Central banks around the world are attempting to use e-cash as a foundation for their Central Bank Digital Currencies (CBDCs). This thesis proposes a protocol for the digital euro, the CBDC of the European Central Bank.

For the past decade, the share of digital payments has increased and the number of cash payments has declined \cite{DNB_2022}. 
With that, the dependency on a connection to an online infrastructure during a transaction (digital payment) has increased, as these digital payment systems require a connection to such an infrastructure to complete the payment. For example, when you pay at a store with a debit or credit card, a connection to your bank is needed to verify whether you have enough balance to pay for the products or services you want to buy. 
Additionally, the balance must be transferred from the payer's account to the one of the payee.
The dependency on an online infrastructure leads to the first main concern that the digital payment system is unusable whenever its online infrastructure cannot be contacted. 
The infrastructure could for example be unreachable in regions with no internet coverage, when the bank's servers are down or during a power outage. 
The latter is increasingly becoming a challenge as the number of power outages has increased for the past years \cite{bhusal2020power} and is expected to increase further \cite{andresen2023understanding, perera2020quantifying}. 
% As a significant share of these power outages is caused by extreme weather events \cite{souto2024identification, schaller2016leading, shield2021major, casey2020power}, Due to climate change, the likelihood and extremity of these weather events have increased \cite{stott2016climate, ebi2021extreme, IPCC2023}, which could cause even more frequent power outages. 

The critical dependency on an online infrastructure for digital payments is not limited to banks, as cryptocurrencies have the same dependency.  
For example, in the case of Bitcoin \cite{nakamoto2008bitcoin}, payments are stored in a data structure that is maintained by all participants that maintain the online infrastructure, a blockchain. This blockchain is used to verify if payments are valid and completed or not. If this blockchain cannot be reached, the receiver cannot verify if he correctly received the payment.

Cryptocurrencies that support offline digital payments, like Zcash \cite{zcash}, still require the spender to prepare the transaction online to be signed later. This requires spenders to know the exact transaction details up-front. These details include the receiver's account number (a "wallet address") and the value of the transaction. This makes the payment method impossible to use for unexpected transactions or transactions where the value cannot be determined up-front. 

The second main concern for digital payment systems is that third parties can observe balance transfers.
Current mainstream digital payment methods offer less privacy than cash payments.
We consider digital payments through banks, cryptocurrencies and e-cash.
Banks have a complete list of all transactions involving the account holders and their balances. In case of a breach, a full list of transactions could be found and potentially abused. 
However, when using cash, the bank only knows who withdrew cash when or deposited cash when.

For most cryptocurrencies, the second type of digital payment system that we consider, payments are stored in a blockchain, using a wallet address as a pseudonym. A wallet address is a unique identifier of a cryptocurrency wallet that is used to retrieve and send cryptocurrencies. The wallet addresses of users may change or remain the same.  
For instance, in Ethereum \cite{ethereum}, users have a fixed wallet address, whereas in Bitcoin \cite{bitcoin} users can change the wallet addresses with every transaction. 
If you know which address belongs to someone, the payment executed with that wallet address can be traced and linked to their identity. 
Even with changing wallet addresses, payments and identities can be traced.
By marking outgoing payments of a wallet address and any subsequent payments from the wallet address that received the payment, flows of payments between wallet addresses can be constructed. 
This flow of payments exposes links between payments and wallet addresses owned by a single identity. The method of marking payments is called taint analysis \cite{tironsakkul2022context}.  

The last digital payment option that we consider is e-cash.
e-cash offers users more privacy, compared to bank payments and cryptocurrencies.
Similar to regular cash, a user must first withdraw e-cash from the bank, before it can be used.
e-cash is represented as a digital token and can be stored on a device. 
At a later stage, the holder uses the token(s) to pay by transferring the tokens to the receiver. 
Finally, the receiver can deposit the tokens at the bank to redeem the value of the tokens. 

The goal of e-cash is to provide properties similar to physical cash. There are three key properties: anonymity, unforgeability and unlinkability. In e-cash, anonymity is said to be provided when an attacker cannot link an identity to a payment \cite{dreier2015Formal}. Unforgeability states that it should be impossible for users to create fake tokens, that appear valid, in the issuer's name. Unlinkability states that it should be impossible to link any payment to a user, even if the user's identity is known.  

In contrast to digital payment systems based on banking infrastructure or cryptocurrencies, some e-cash protocols also support offline payments. 
In a fully offline e-cash scheme, no bank, ledger, or other third party is involved in the payment between the spender and the receiver. 
Because e-cash provides fully offline privacy-preserving payments, it satisfies both main concerns (online infrastructure and exposed payments) of the current digital payment infrastructure. Therefore we explore it further in this work.

Many central banks have expressed their interest in e-cash and some central banks are already using e-cash as a digital version of their currency. 
These digital versions of currencies backed by a central bank are named Central Bank Digital Currencies (CBDCs). 
In December 2023, 130 countries, contributing to 98\% of the global GDP, have expressed their interest in a CBDC, are researching and developing it, or have a CBDC in circulation \cite{Atlantic_Council_2024}. 
A survey from the International Monetary Fund (IMF) \cite{imfTakingOffline} found that most CBDCs in development can only be used online. This implies that they have the same critical dependency on an online infrastructure as most other digital payment systems. Moreover, compared to online CBDCs, offline-capable CBDCs promote financial inclusion, have lower transaction costs and improve the user experience \cite{michalopoulos2024}. 

The CBDCs that can be used offline typically rely on tamper-resistant hardware to maintain the integrity of the CBDCs stored on a device \cite{imfTakingOffline}. However, as Liu et al. \cite{liu2021understanding} and Lee et al. \cite{lee2017hacking} have shown, even the current state-of-the-art hardware can be breached and thus fails to achieve integrity through tamper-resistance. Relying solely on hardware to prevent users from misusing the e-cash, is thus not sufficient.
Therefore, the CBDC should guarantee integrity through software. 
This can be done using cryptographic protocols, such as proofs and signatures, to maintain the protocol's integrity, as is done in \cite{brands1994untraceable, liu2005recoverable, eslami2011new, baseri2013secure, fan2014date, baldimtsi2015anonymous, bauer2021transferable, 9650670}. 
This approach makes it possible to detect fraud retrospectively and to revoke malicious users their anonymity.

The European Central Bank (ECB) is currently in the 'preparation stage' of designing the (offline) digital euro \cite{EuropeanCentralBank2023}. 
Two of the main design goals of the digital euro are protecting privacy as much as possible and support for offline transactions \cite{EuropeanCentralBank2023cont}. This thesis will look at a software implementation of the digital euro, that fulfils the two design goals set by the ECB. 

Following the approach of the Office of Science and Technology Policy \cite{technicalDesignChoices} for a digital U.S. Dollar and the recommendation of the European Data Protection Supervisor \cite{techdispatch2023}, our implementation will use digital pseudonyms and zero-knowledge proofs.
A zero-knowledge proof is a type of cryptographic proof used to prove knowledge of a secret, without revealing the secret. 
In our proposed protocol for the digital euro, zero-knowledge proofs are used as payment proofs. 
This way other users and banks can verify the proofs and the transaction without being able to identify the previous spenders when they receive a token. To further protect the users' privacy, participants in our system operate under digital pseudonyms. 

This thesis proposes a design for the payment protocol of the digital euro that supports fully offline payments, whilst protecting the users' privacy. 
The protocol relies on zero-knowledge proofs and signatures to protect its integrity and to provide retrospective fraud detection. 
To demonstrate the protocol's correctness and functionality, the protocol is also implemented in an open-source minimum viable product. More specifically our contributions are as follows:

\begin{enumerate}
    \item We propose a payment protocol that supports fully offline privacy-preserving payments and could be used by the ECB as implementation for the digital euro. The protocol uses zero-knowledge proofs and signatures providing privacy and retrospective fraud detection (Section \ref{PaymentProtocol}).
    \item We provide an open-source minimum viable product of the protocol to demonstrate the correctness and functionality (Section \ref{ImplementationSection}).
\end{enumerate}
\section{Problem description}
There are three major problems regarding e-cash: balancing privacy and fraud prevention, double spending and token transferability.
The first major problem is the trade-off between privacy and fraud prevention.
Fraud can be trivially detected and prevented by making the transactions traceable and fully linkable to users. 
This would, however, require all participants in the system to give up their privacy and reveal sensitive information regarding their spending behaviour. 
When the architecture requires traceability and full linkability, the e-cash has the same concern regarding privacy preservation as banking payments and payments done with cryptocurrencies.

The second problem, double-spending \cite{baldimtsi2015anonymous, bauer2021transferable}, results from having too much privacy. Receivers of an e-cash token cannot check if the same token was used in an earlier payment. The spender could thus copy a single token and spend the copy at multiple places. In an analogy to cash, this would mean copying a banknote and spending that forged banknote at different locations. In a fully anonymous setting, a third party cannot link an identity to the fraudulent payment. This makes the e-cash scheme more vulnerable when too much privacy is given to the users.  

Currently available solutions to the problem of double-spending, show a necessary restriction to our system's implementation. As mentioned earlier, the secured hardware can be breached, thus the implementation is restricted to retrospective double-spending detection and anonymity revocation. As an additional restriction, it should be possible to complete fully offline transactions. This means that the double spending detection cannot rely on third parties involved in the transaction.

The third problem is token transferability. Token transferability is a property of e-cash that allows users to reuse a token they received in an earlier payment. This means that a token can have multiple holders, and be used in multiple transactions, before being deposited. To satisfy unlinkability, token holders should not be able to find any information regarding the identity of previous holders in the token. 

Combining token transferability with anonymity revocation introduces a new challenge as tokens can have multiple holders. To find the correct identity that committed double spending, details of every transaction must be included with the token. This is because the transaction details include the identity of the spender. As a result, the size of the token grows with every transaction \cite{chaum1992transferred}. 
\section{Related work}
Since the introduction of blind signatures in 1983 by Chaum \cite{chaum1983blind} and the first offline e-cash protocol by Brands \cite{brands1994untraceable}, there has been little (recent) research on offline e-cash schemes that do not rely on hardware or trusted software to prevent double-spending. 
Relying on such hard- or software to avoid double-spending, such as is done in \cite{juang2005practical,juang2010ro, sakalauskas2018simple, luo2018offline, hong2022solution}, is an unrealistic assumption and breaking this hard- or software's integrity would also invalidate the entire e-cash scheme.

The other method to prevent double-spending is to rely on cryptographic principles to detect double spending and revoke the anonymity of the malicious user. 
This occurs when the tokens are deposited at the bank \cite{brands1994untraceable, liu2005recoverable, eslami2011new, baseri2013secure, fan2014date, baldimtsi2015anonymous, bauer2021transferable, 9650670}. Whenever the bank detects two tokens with the same identifier, the transaction details can be used to revoke the identity of the double-spender.

A subset of the protocols that rely on cryptography is unpractical as it allows shops to mint tokens (\cite{batten2019off}) or nonfunctional in a fully offline scenario (\cite{osipkov2007combating}), especially when intended to be used as a basis of a CBDC (Section \ref{Usability}).
Moreover, other research regarding offline e-cash introduces functionality (token expiration \cite{eslami2011new, fan2014date, baseri2013secure}), which does not solve and potentially worsens the problem it was intended to solve (Section \ref{TokenExpiriation}). 

To fully benefit from the offline functionality, token transferability (Section \ref{Transferability}) is highly desirable. With transferable tokens, it is not required to withdraw and deposit a token for each payment. Having the possibility to spend tokens multiple times without the bank's involvement would thus reduce the communication costs between the bank and the users \cite{baldimtsi2015anonymous}.
A prototype of an offline transferable digital euro, EuroToken, was proposed earlier \cite{blokzijl2021eurotoken, koning2023performance} (Section \ref{EuroToken}). 
However, this proposed scheme is fully traceable and offers little privacy to the users and thus conflicts with one of the main design goals (preserving privacy) set by the European Central Bank \cite{EuropeanCentralBank2023cont}.

\subsection{Real world (un)useabilty}
\label{Usability}
Besides the integrity of the e-cash scheme, its useability must be considered. 
Some proposed e-cash schemes make assumptions or functionality that are infeasible or undesirable. 
For instance,  Osipkov et al. \cite{osipkov2007combating} claim to prevent double-spending without trusted hard- or software in an offline setting. 
However, they make use of the assumption that the merchants (receivers) have a functional peer-to-peer network. 
This scenario, where the network is partially offline, does not offer a solution to pay in areas without network coverage. Furthermore, the peer-to-peer network does not satisfy our fully offline constraint.

Another example is the scheme proposed by Batten et al. \cite{batten2019off}, which provides spare change by giving reputable shops, such as Target, the authority to mint cash. Given that our protocol is intended to be used for a CBDC, it is undesirable that shops can mint tokens. 

\subsection{Expiring e-cash}
\label{TokenExpiriation}
Eslami and Talebi \cite{eslami2011new} introduced expiring e-cash by attaching an expiration date to the e-cash description. 
This scheme was later improved by \cite{baseri2013secure} and \cite{fan2014date}. 
The main reason for this expiration date is a storage reduction for the bank. 

The question remains if schemes, that support token expiration, solve the problem of storage required to detect double-spending. 
The option to recover expired e-cash will not lead to a decrease in transactions, given that users withdraw e-cash intending to spend it. Whether or not the e-cash has an expiration date, users would need to pay for the same amount of goods or services, the same number of times.
Therefore, the number of deposits does not change. 
This means that the storage size needed to store all deposited tokens will not be reduced by adding an expiration date. 
Tokens that have been deposited and expired after can not be removed from the storage, because they are needed to check if a token has been spent when it is sent for exchange, as is designed in \cite{eslami2011new, baseri2013secure, fan2014date}.
Furthermore, by offering an exchange service for expired tokens, the bank should store the exchanged tokens leading to a larger required storage to detect double-spending. Given that adding an expiration date to tokens increases the storage size required, our protocol does not include token expiration.

\subsection{Transferable e-cash}
\label{Transferability}
Transferability is a highly desired property that e-cash should have. 
Transferable e-cash makes it possible to spend the e-cash received by other users without depositing and withdrawing new e-cash first. 
This reduces the dependency on reaching the bank even further and does not limit the number of transactions to the number of withdrawals. 

The downside of transferability is that it requires the e-cash to grow in size with each transaction. 
This is because storing information about every transaction is needed to reveal the double spender's identity \cite{chaum1992transferred}. 

Sarkar \cite{sarkar2013multiple} tried to achieve this property using bitwise XORs. 
However, the protocol uses an unspecified distributive operator over XOR to detect double-spending \cite{barguil2015efficient}. 
Furthermore, Barguil \cite{barguil2015efficient} also proves that the security claims made by Sarkar do not hold. 

Baldimtsi et al. \cite{baldimtsi2015anonymous} proposed a transferable e-cash scheme using malleable signatures. 
This type of signature is used to sign transactions whilst keeping the bank's signature valid. 
Double-spending is detected by making use of cryptographic tags. For each transaction, a \textit{double spending (DS) tag} and a \textit{serial number (SN) tag} are created and added to the token during a transaction. The SN tag is used in the next transactions as the identifier of the token and is computed using the DS tag of the previous transaction. The DS tag is constructed using the SN tag used in the transaction and is used by the bank to detect double-spending. A bank could detect double spending upon deposit when it receives two tokens, having the same SN tag with a different DS tag.   

The protocol of Baldimtsi et al. is improved by Bauer et al. \cite{bauer2021transferable} by replacing the inefficient malleable signatures with a commit-and-proof scheme. 
With this scheme, the tags to detect double spending are also randomized in each transaction.
Our protocol has a similar approach with a commit-and-proof scheme. However rather than using two tags, proofs are added to the token to detect double-spending.  

Jianbing et al. \cite{9650670} tried to take the transferability one step further by proposing a transferable e-cash scheme that allows the receiver to be anonymous and thus provides dual anonymity. 
However, they used a different definition of transferable, as the protocol requires users to contact the bank to re-randomize a received token after each transaction. Having to re-randomize a token at the bank before you can reuse it, removes the benefits (reducing the dependency on the bank) of transferable e-cash.

\subsection{EuroToken}
\label{EuroToken}
Blokzijl \cite{blokzijl2021eurotoken} and Koning \cite{koning2023performance} of the Tribler Lab \footnote{https://github.com/Tribler/tribler/wiki\#current-items-under-active-development} and the Delft University of Technology did earlier work regarding a CBDC for the ECB, named EuroToken. 
This work was done in collaboration with the Nederlandsche Bank. 
This thesis serves as a continuation of their research.

In the scheme of Blokzijl and Koning, the bank mints a token by defining a serial number, a face value and a nonce. 
Upon withdrawal, the bank sends the user the minted token, a tuple of the receiver's public key and a signature of the bank on the minted token and the receiver's public key.

The signature tuple is the start of a chain of proofs of ownership. 
This chain of ownership is sent with the token and is extended with each transaction. 
As the bank's signature includes the withdrawer's public key, the withdrawer can prove he owns the token. 
When the user spends the token, the user will send the token and extend the chain of ownership with a tuple of the receiver's public key and a signature, singing the previous proof of ownership and the recipient's public key. 
The deposit of the token is similar to a transaction between users. 
However, now the bank is the receiver of the token. After the bank has received the token it can verify the proof chain and check for double-spending.

Token holders can verify the chain of ownership after $k$ transactions starting from the bank's signature. 
This signature can be used to find the public key of the first receiver. 
The found public key can then be used to validate the next proof and to find the next recipient's public key. 
After $k$ transactions the last found public key maps to the current holder of the token. 

The bank can detect double spending upon deposit of the tokens. 
Whenever the bank has received two tokens with the same first proof double spending must have occurred. 
The bank can then compare the chain of proofs of ownership to find the double spender. 
After some $i$ proofs there must be two proofs where proof $i + 1$ from the first chain differs from proof $i + 1$ from the second chain. 
This implies that proof $i$ is used in two transactions and thus doubly spent. 
The identity of the double spender can then easily be found, as that is the receiver's public key used to create proof $i$.

The problem with this proposal is that it offers no privacy and the token's history is fully traceable. 
Whenever someone receives a token, all the public keys of the previous holders can be found. 
Malicious people who know which public keys map to which identity could use and abuse that information to obtain sensitive personal information when receiving a token. 

Privacy is an important factor in why people use cash for payments \cite{RoleOfCash}. 
The current implementation of EuroToken offers less privacy than the online payment infrastructure of banks. 
This combined will have a detrimental effect on the adoption rate of the CBDC, as the option of paying offline will come at the cost of user privacy. 
Moreover, the provided protocol does not align with the main design goal of the ECB, namely privacy protection \cite{EuropeanCentralBank2023KeyObjectives}.
\section{Security assumptions}
The protocol proposed by this thesis relies mainly on two security assumptions to guarantee unforgeability and anonymity. 

These assumptions are the \textit{Discrete logarithm problem} and the \textit{Computational Diffie-Hellman} assumption.
The Discrete logarithm problem states that given a finite cyclic group $G$, generator $\langle g \rangle $ of $G$ and $h \in G$, it is hard to find an integer $a$, such that $g^a = h$. 
This hardness will be used to create unforgeable signatures and proofs of ownership.

The Computational Diffie-Hellman assumption states that given a finite cyclic group $G$, generator $\langle g \rangle $ of $G$, $g^a$ and $g^b$, it is computationally hard to compute $g^{ab}$, without knowing the values of $a$ and $b$. 
This assumption is used to verify knowledge of the private key and as a basis for the security of bilinear pairing cryptography.

\section{Signatures and Groth-Sahai proofs}
Our system has two main cryptographic components, blind signatures and Groth-Sahai proofs. 
The blind signature is applied to prevent the bank from linking the withdrawn digital euro to the first holder. 
The Groth-Sahai proofs are used to create a zero-knowledge proof of a transaction, to provide anonymity between transactions. These proofs are constructed with bilinear pairings.

\subsection{Blind signatures}
\label{sectionBlindSignatures}
Chaum \cite{chaum1983blind} first introduced blind signatures in 1983. 
A blind signature scheme can be used to obtain a valid signature on a message $M$, without the signer knowing the exact content of $M$. 
This makes it possible for e-cash to have a valid signature of a bank for an unknown token. 
When this token is deposited later, the bank cannot recognize which user has withdrawn the token. 
This makes it impossible for the bank to link the user who withdrew the token to the user who deposited it, proving anonymity. 
In this thesis, an implementation of a hash-based blind Schnorr signature (BSS) is used. 

As the (blinded) Schnorr signature scheme is based on groups, there should be a group $g$ with order $q$ known by both the client (a user) and the signing party (the bank). 
Furthermore, the signing party chooses a random private key $x \in_R\mathbb{Z}_q^*$ and shares the public key $y = g^x$ and a cryptographic hash function $H$ with the clients. 
A BSS on message $M$ can then be obtained as follows: 

\begin{enumerate}
    \item The signing party chooses a random $k \in_R\mathbb{Z}_q^*$ and sends $r = g^k$ to the client.
    \item The client picks random blinding factors $\alpha, \beta \in_R\mathbb{Z}_q^*$ and calculates $r'$ as $r' = rg^{-\alpha}y^{-\beta}$.
    \item With that the client computes the challenge $c$ for message $M$: $c = H(r' || M) \mod q$, and sends blinded challenge $c' = c + \beta$ to the signing party.
    \item The signing party then signs the blinded message as: $\sigma' = k - c'x$ and returns $\sigma'$.
    \item To obtain the signature on message $M$ the client computes: $\sigma = \sigma' - \alpha$. The Schnorr signature is then defined as $(\sigma, c)$ 
    \item Other parties can verify the validity of the signature on message $M$ by computing $r_v = g^\sigma y^c$ and checking: $c \stackrel{?}{=} H(r_v||M)$.    
\end{enumerate}

A mathematical protocol description of the BSS protocol can be found in Figure \ref{tab:blindsignature}.

\begin{table}[h]
\normalsize
    \centering
    \begin{tabular}{l c r}
        Client & & Signing party \\
         & & $k \in_R\mathbb{Z}_q^*$ \\
         & & $r \leftarrow g^k$ \\
         & $\xleftarrow{r}$ & \\
         $\alpha, \beta \in _R\mathbb{Z}_q^*$ & & \\
         $r' \leftarrow rg^{-\alpha}y^{-\beta}$ & & \\
         $c \leftarrow H(r' ||M)$ & & \\
         $c' \leftarrow c + \beta$ & & \\
         & $\xrightarrow{\text{$c$}}$ & \\
         & & $\sigma' \leftarrow  k - c'x$ \\
         & $\xleftarrow{\sigma'}$ & \\
         $\sigma \leftarrow \sigma' - \alpha$ & & \\
    \end{tabular}
    
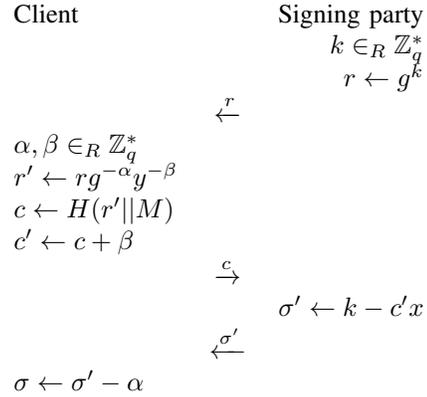
\captionof{figure}{ Blind Schnorr signature protocol to obtain a blind signature ($\sigma$, $c)$ on message $M$}
    \label{tab:blindsignature}
\end{table}

The blind signature is done over the hash of the message to prevent malicious clients from creating more valid signatures from an earlier received signature. 
Due to the multiplicative homomorphic property, malicious clients could also compute valid signatures on multiples of message $M$ without the hash.

Given that the hash function is collision-resistant, it is hard for a malicious client to find the message corresponding to the malled signature. 
Therefore it is impossible to create more valid signatures, based on an earlier received signature.

\subsection{Bilinear pairings}
A bilinear map $e$ is an operation that takes two elements from, potentially, different elliptic curve groups of order $p$ and maps them to an element of a third group, the target group. 
More formally, given source groups $G$, $H$ and target group $G_T$, a bilinear map is denoted as: 
\begin{align*}
    e:G \times H \rightarrow G_T
\end{align*}

Additionally, the pairing must satisfy the following three properties:
\begin{itemize}
    \item \textbf{Bilinearity:} For all items $P, Q \in G$ and $R, S\in H$, the following holds:  \\
    \begin{align*}
        e(P + Q, R) = e(P,R) \cdot e(Q,R) \\
        e(P, R + S) = e(P,R) \cdot e(P,S)
    \end{align*}
    Moreover, given generators $g, h$ such that $G = \langle g \rangle$ and $ H=\langle h \rangle$, for all $a,b \in \mathbb{Z_p}$ the following holds:
    \begin{align*}
        e(g^a, h^b) = e(g,h)^{ab} \\
    \end{align*}
    \item \textbf{Non-degeneracy:} $e(P, R) \neq 1$.
    \item \textbf{Efficient computability:} There must be an efficient method to calculate the pairing efficiently.
\end{itemize} 
\vspace{\baselineskip}
An extended bilinear map $E$ is a mapping of two elements of $G$ and two elements of $H$ to four elements of $G_T$:
\begin{align*}
    E:G^2 \times H^2 \rightarrow G_T^4
\end{align*}
As an example, given $g_1, g_2 \in G$ and $h_1, h_2 \in H$:
\begin{equation}
\label{ExpandedMaps}
    E\begin{pmatrix}\begin{pmatrix}g_1 \\ g_2 \end{pmatrix}, \begin{pmatrix}h_1 & h_2 \end{pmatrix}\end{pmatrix} = \begin{pmatrix} e(g_1,h_1) & e(g_1,h_2) \\  e(g_2,h_1) & e(g_2,h_2) \end{pmatrix}
\end{equation}
Similarly to regular bilinear maps, the extended bilinear maps are also bilinear, using entry-wise product operations for the vectors and matrices. Given $g_1, g_2, g_3, g_4 \in G$ and $h_1, h_2 \in H$:
\begin{align*}
\resizebox{0.5\textwidth}{!}{$
    E\begin{pmatrix}\begin{pmatrix}g_1 \\ g_2 \end{pmatrix}\begin{pmatrix}g_3 \\ g_4 \end{pmatrix}, \begin{pmatrix}h_1 & h_2 \end{pmatrix}\end{pmatrix} = 
    E\begin{pmatrix}\begin{pmatrix}g_1 \\ g_2 \end{pmatrix}, \begin{pmatrix}h_1 & h_2 \end{pmatrix}\end{pmatrix} E\begin{pmatrix}\begin{pmatrix}g_3 \\ g_4 \end{pmatrix}, \begin{pmatrix}h_1 & h_2 \end{pmatrix}\end{pmatrix}
    $}
\end{align*}

\subsection{Groth-Sahai proofs}
\label{Groth-Sahai Proofs}
In 2008, Groth and Sahai \cite{groth2008efficient} presented a proof framework that can be used to efficiently create non-interactive zero-knowledge (NIZK) proofs and non-interactive witness-indistinguishable (NIWI) proofs. 
Before this, NIZK proofs were inefficient and thus not useable. 
The Groth-Sahai (GS) proofs are designed to prove statements in pairing-based equations. 

As a setup, a (trusted) party must publish a bilinear pairing description and a Common Reference String (CRS). 

The bilinear pairing description is defined as:
\begin{align*}
    (G_1, G_2, G_T, e, g_1, g_2)
\end{align*}
in which $G_1$ and $G_2$ are two bilinear groups. 
These groups have a mapping $e$ to target group $G_T$. 
$g_1$ and $g_2$ are generators of respectively $G_1$ and $G_2$. 
When $G_1 \equiv G_2$ the pairing is symmetric and if $G_1 \neq G_2$ the pairing is asymmetric. Functionally, both types of pairings would work for our protocol. Symmetric pairings allow for an easier protocol description \cite{chatterjee2010efficiency}. However, asymmetric implementations tend to perform better on higher security levels.

The CRS is constructed with two pairs of four random group elements, four from $G_1$ and four from $G_2$ and is defined as:
\begin{align*}
    CRS = (g, u, g', u', h, v, h', v')
\end{align*}

Depending on the structure of the GS proofs, the CRS can be used in a trapdoor function. 
In some proof structures, this will reveal the input and can thus be used to find the values used to create the committed value (the value to be proven). In our protocol (Section \ref{ImplementationSection}) this is used to find the public key of the spender.
The setup can be done with public randomness and multiple parties to fully remove the trust needed in a (central) party.     
Each proof consists of three parts, namely the target $T$, the commitment values $c_1, c_2, d_1, d_2$ and proof elements $\theta_1, \theta_2, \pi_1, \pi_2$. 
The target represents the value that the prover wants to prove. 
The commitment values are used to randomized encryptions of values with which the proof is constructed. 
Elements from $G_1$ are encrypted in $c_1$ and $c_2$, whereas elements from $G_2$ are encrypted in $d_1$ and $d_2$. 
Lastly, the proof elements are used to derandomize the commitment values without revealing the exact values.

In our protocol, the implementation of the Groth-Sahai proofs is as follows. 
The equation to prove is $e(X, Y) = T$ in which $X \in G_1$ and $Y \in G_2$ and $T$ is the target of the proof. 
The commitment values are randomized with values $r, s \in Z_p$, and computed as:\\

\begin{table}[H]
    \centering
    \normalsize
    \begin{tabular}{l l}
$c_1 = g_1^r$   & $d_1 = g_2^s$\\
$c_2 = u^rX$ & $d_2 = v^sY$\\
    \end{tabular}
\end{table}
\noindent
%$c_1 = g_1^r$ \\
%$c_2 = u^rX$ \\
%$d_1 = g_2^s$ \\
%$d_2 = v^sY$ \\
\noindent
The prover now picks a random value $t \in Z_p$ and computes the proof elements as: \\
%$\pi_1 = d_1^rg_2^t$ \\
%$\pi_2 = d_2^rv^t$ \\
%$\theta_1 = g_1^{-t}$ \\
%$\theta_2 = Xu^{-t}$ \\
\begin{table}[H]
    \centering
        \normalsize
    \begin{tabular}{l l}
    $\theta_1 = g_1^{-t}$ & $\pi_1 = d_1^rg_2^t$ \\
    $\theta_2 = X^su^{-t}$ & $\pi_2 = d_2^rv^t $\\
    \end{tabular}
\end{table}
\noindent
The full proof is now defined as $(c_1, c_2, d_1, d_2, \pi_1, \pi_2, \theta_1, \theta_2)$ and can be verified by others with Equation \ref{GSVerification}. 

\begin{equation}
\label{GSVerification}
\resizebox{0.5\textwidth}{!}{$
    E\begin{pmatrix}\begin{pmatrix}c_1 \\ c_2 \end{pmatrix}, \begin{pmatrix}d_1,d_2 \end{pmatrix}\end{pmatrix} \stackrel{?}{=} 
    E\begin{pmatrix}\begin{pmatrix}g_1 \\ u \end{pmatrix}, \begin{pmatrix}\pi_1, \pi_2\end{pmatrix}\end{pmatrix} 
    E\begin{pmatrix}\begin{pmatrix}\theta_1 \\ \theta_2 \end{pmatrix}, \begin{pmatrix}g_2, v\end{pmatrix}\end{pmatrix} 
    \begin{pmatrix}1 & 1 \\ 1 & T \end{pmatrix} 
    $}
\end{equation}
The verification can be done elementwise after expanding the extended bilinear maps as in Equation \ref{ExpandedMaps}. 
For example, to verify $e(c_1,d_1)$, the following must hold:
\begin{align*}
    e(c_1,d_1) \stackrel{?}{=} e(g_1, \pi_1)\cdot e(\theta_1,g_2)\cdot 1
\end{align*}

If someone knows the exponents used to create $u$ and $v$ from the CRS, one could find the committed values of $X$ and $Y$.
Let $u = g_1^\alpha$ and $v = g_2^\beta$, the committed values can be retrieved with the equations \ref{Knowledge extractor}a and \ref{Knowledge extractor}b. \\
\begin{subequations}
\label{Knowledge extractor}
\begin{align}
X &= c_1^{-\alpha}c_2\\
Y &= d_1^{-\beta}d_2
\end{align}
\end{subequations}
%In the second implementation of GS-proof (TYPE-II), the target is defined as $xy = T$, in which %$x \in 
%Z_p$ and $y \in Z_p$. With random values $r, s \in Z_p$ the commitment values are calculated as: %\\
%\begin{table}[H]
%    \centering
%    \begin{tabular}{l l}
%    $c_1 = g^r(g')^x$ & $d_1 = h^s(h')^y$\\
%    $c_2 = u^r(u'g)^x$ & $d_2 = v^s(v'g)^x$\\
%    \end{tabular}
%\end{table}
%
%%$c_1 = g^r(g')^x$ \\
%%$c_2 = u^r(u'g)^x$ \\
%%$d_1 = h^s(h')^y$ \\
%%$d_2 = v^s(v'g)^x$ \\
%
%\noindent
%The prover now picks a random value $t \in Z_p$ and computes the proof elements as: \\
%
%\begin{table}[H]
%    \centering
%    \begin{tabular}{l l}
%    $\pi_1 = d_1^rh^t$ & $\pi_2 = d_2^rv^t$\\
%    $\theta_1 = g'^{xs}g^{-t}$ & $\theta_2 = (u'g)^{xs}u^{-t}$\\
%    \end{tabular}
%\end{table}
%
%%$\pi_1 = d_1^rh^t$ \\
%%$\pi_2 = d_2^rv^t$ \\
%%$\theta_1 = g'^{xs}g^{-t}$ \\
%%$\theta_2 = (u'g)^{xs}u^{-t}$ \\
%\noindent
%Similar to the first type of GS-proof, the full proof is defined as $(c_1, c_2, d_1, d_2, \pi_1, %\pi_2, \theta_1, \theta_2)$ and can be verified by others with equation \ref{GSVerification}. %However, if someone knows the exponents used in the CRS, it is impossible to find the values of %$x$ and $y$.%
\section{Digital euro protocol}
\label{PaymentProtocol}
Our protocol is divided into four phases: initialization, withdrawal, transactions and deposit. The initialization phase is executed only once by a trusted third party (TTP), the bank and the users. The other three phases are related to the cycle of a single digital euro. 

\subsection{Initialization}
In the initialization phase, the TTP responsible for managing identification publishes a bilinear pairing description and a common reference string (CRS), as found in Section \ref{Groth-Sahai Proofs}. The exponents used to generate the group elements are stored for later use by the TTP but remain private. The participants in the protocol will use the bilinear pairing description and CRS. 

Every participant has to register at the TTP as well. Upon registering the user picks a random private key $x$, calculates the public key $X = g_1^x$ and registers $X$ at the TTP. The user can register at a bank with the public key, certified by the TTP. The bank can use this public key to keep track of the user's balance.

\subsection{Withdrawal}
At the start of the withdrawal phase, the user can prove his identity to the bank in the same way as during the initialization phase. After that, the BSS protocol (Section \ref{sectionBlindSignatures}) is used with the generator $g_1$ of order $p$ of the bilinear group description provided by the TTP. 

The message to be signed consists of the serial number and a random group element. The withdrawer can generate a serial number randomly. This serial number is used for tracking but does not need to be unique however the combination with the random element should be unique. For the random group element, the user picks a value $t \in _R\mathbb{Z}_p^*$ and computes $\theta_1 = g_1^{-t}$. This $t$ will be later used in a transaction to demonstrate knowledge of randomization. The serial number and $\theta_1$ can then be converted to bytes and concatenated to be blindly signed by the bank. When the protocol is completed the digital euro is described as: 
\begin{align*}
    (SN, \theta_1, \sigma, GS)
\end{align*}
in which, SN is the serial number of the digital euro, $\theta_1$, $\sigma$ is the blind signature of the bank on $SN$ and $\theta_1$ and $GS$ is an ordered list of Groth-Sahai proofs of previous transactions. Upon withdrawal $GS$ is empty.

\subsection{Transactions}
\label{TransactionsSection}
Every transaction a digital euro has undergone must be stored with the euro to combat double-spending. To find the user that double-spent a euro, the details of the malicious transaction must be known to retrieve the identity of the double-spender, as shown in \cite{chaum1992transferred}. This scheme stores the required information as a GS proof. By storing the information in a zero-knowledge proof, participants in later transactions, or the bank, cannot deduce any information related to the transaction from the proof. They can, however, verify if the proofs and thus the transactions are valid. During a transaction, the spender and the receiver collaborate to create a GS proof, which is stored with the digital euro.

To start a transaction the receiver generates a random $t$ and sends the randomization elements $g_2^t$, $v^t$, $g_1^{-t}$ and $u^{-t}$ to the spender, whilst keeping $t$ secret. This prevents the spender from deciding on all randomness and trying to obfuscate double-spending by using the same randomness for two transactions with the same digital euro. When the same randomization is used in two transactions with the same token, the proofs for those transactions are the same, hiding the double-spending. The spender will use these randomization elements given by the receiver when creating the GS proof for the transaction.

The target of the proof for transaction $i$, $T_i$, depends on whether the digital euro is spent earlier. When the euro has not been spent before, the target is $T_0 = e(g_1,g_2)^\sigma$. Otherwise, after $i$ transactions the target can be computed as $T_i = e(g_1,g_2)^{T_{i-1}}$. This way, the targets of the proofs can be used to describe a chain of transactions, in which the current proof links to the previous proof.

Let $k = \sigma$ if $i = 0$, and $k = T_{i-1}$ otherwise. With $k$, the spender can compute $y=\frac{k}{x}$ and $Y = g_2^y$, in which $x$ is the spender's private key. The spender can now use the GS proof, to prove $e(g_1^x, g_2^y)$. Note that $g_1^x$ is equal to the spender's public key. Additionally, due to the property of bilinearity, $e(g_1^x, g_2^y) = e(g_1,g_2)^{xy} = e(g_1, g_2)^{k} = T_i$. 

The value of $s$ in the proof is set to the inverse of $-t_{prev}$, the $t$ used in the previous transaction to provide the randomization elements. This implies that the spender must know the value of $t$ used during the last transaction and cannot generate a valid proof if he does not. For the first transaction, no $t_{prev}$ is available. However, the spender in the first transaction can use the $t$ used in the withdrawal phase as he is the withdrawer.

To prevent the receiver from creating valid proofs by changing the values of t after the transaction, the spender also computes an additional signature. This is a Schnorr signature constructed with signing key $r$ used to create GS proof and signs the value of $g_1^{-t}$. This signature only has to be shown in the next transaction. The next receiver can verify this signature as the decryption key $g^r$ is provided in the GS of the current transaction as $c_1$.

The spender sends the values of $v^s$ and $Y$ together with the proof elements, the signature received in the previous transaction and the signature of the current transaction to the receiver. With these, the receiver can verify the proof, if $e(X, Y) = T$, check if $d_2$ is constructed correctly and verify the signatures. 

Additionally, the receiver must check if the previous proofs included with the digital euro are correct and verify the links between the proofs. Given the proofs for transaction $i - 1 = j$ and $i$ as:  
\begin{align*}
    (c_{1j}, c_{2j}, d_{1j}, d_{2j}, \theta_{1j}, \theta_{2j},\pi_{1j}, \pi_{2j}, T_{j})
\end{align*}
and
\begin{align*}
    (c_{1i}, c_{2i}, d_{1i}, d_{2i}, \theta_{1i}, \theta_{2i},\pi_{1i}, \pi_{2i}, T_{i})
\end{align*}
the equations \ref{LinkVerficationEquations}a and \ref{LinkVerficationEquations}b must hold:
\begin{subequations}
\label{LinkVerficationEquations}
\begin{align}
T_i &\stackrel{?}{=} e(g_1, g_2)^{T_j} \\
e(\theta_{1j}, d_{1i} ) &\stackrel{?}{=} e(g_1, g_2)^1
\end{align}
\end{subequations}

Equation \ref{LinkVerficationEquations}b must hold to verify that every spender knew the randomization element $t$ in the previous transaction. As $g_1$ and $g_2$ are part of the bilinear pairing description and thus constant, the equation expands to $e(g_{1}^{-t_j}, g_{2}^{s_i})$, which is equal to $e(g_1, g_2)^{-t_js_i}$. For the transaction to be valid $s$ should be the inverse of $-t$ of the previous transaction, implying that $-t_js_i = 1$. This results in the verification form $e(g_1, g_2)^1$.

\subsection{Deposit}
\label{DepositSubsection}
A digital euro can be deposited to the bank in the same way as a digital euro is transferred between users in Section \ref{TransactionsSection}. However, in this case, the bank is the receiver. As the user that wants to deposit the euro has to share their public key, the bank knows to which account the balance should be added. The bank also checks if the digital euro is doubly spent or not.

\subsection{Double spending detection}
The bank detects double spending when two digital euros $DE$ and $DE'$ with the same signature $\sigma$ are deposited. There are two possible scenarios in this case. 

The first trivial case is when $GS$ of $DE$ equals $GS$ of $DE'$, excluding the last proof created in Section \ref{DepositSubsection}. This occurs if, and only if, the same user tries to deposit the same digital euro twice. To deposit the euro the user must identify himself, therefore the identity of the double spender is revealed. 

In the second scenario, when  $GS$ of $DE$ does not equal $GS$ of $DE'$, the bank must take additional actions to reveal the identity of the double spender. Given that the two lists of proofs are different, there must be an index $i$, such that $GS_{DE}[i] \neq GS_{DE'}[i]$. Assuming that the odds that the double spender retrieved the randomization elements generated by the same $t$ are extremely unlikely, the proofs have, at least, different values for the $\theta_1$ and $\theta_2$ proof elements.

The bank can then send both proofs to the TTP. The TTP can extract the public key $X$ with Equation \ref{Knowledge extractor}a, for both proofs and check if $X$ is the same for both proofs sent by the bank. If they are the same, the TTP can retrieve the legal identity, registered with this public key, and return it to the bank. Otherwise, this transaction is no occurrence of double-spending. This could for example occur when the double spender did receive the same randomization parameters. 

\subsection{Efficiency analysis}
As mentioned earlier, the size of the digital euro must grow to detect double spending and revoke the anonymity of the double-spender. As seen in Section \ref{TransactionsSection}, every transaction included in the digital euro is defined in a GS-proof. This means that the size of the digital euro grows with 8 or 9 group elements for each transaction. The number of group elements depends on whether the value of $T$ is explicitly included in the proofs. Given that the target $T$ can be calculated from the proof elements of the previous proof, it can be omitted for size optimizations. This means that the size of the digital euro after $n$ transactions (using a symmetric pairing) can be computed as:

\begin{equation}
\label{sizeequation}
    size = |SN| + |G| + |\sigma| + n \cdot 9|G|
\end{equation}

in which $|SN|$ denotes the size of the serial number, $|\sigma|$ the size of the signature of the bank and $|G|$ the size of a group element.

\section{Implementation and experiments}
\label{ImplementationSection}
The described protocol is implemented in Kotlin as a proof of concept. 
The open-source implementation can be found on GitHub \footnote{https://github.com/LeonKempen/trustchain-superapp/tree/master/offlineeuro}. 
The Java Pairing Based Cryptography (JPBC) library \cite{jpbc} is used for group and bilinear map operations. 
As this is a proof of concept, it is not a fully implemented financial system and users can freely withdraw and deposit digital euros without affecting their balances.
This does not affect the results as the balance of the user's bank account does not influence the digital euro protocol.
This prototype was used to test the protocol for correctness, growth size and verification performance. 
The tests were performed on a desktop with an Intel Core i5-4590 (3.30GHz) processor and 8 GB of RAM.

\subsection{Prototype}
\begin{figure*}[ht]
  \begin{subfigure}{0.31\textwidth}
    \fbox{\includegraphics[width=\linewidth, height=10cm]{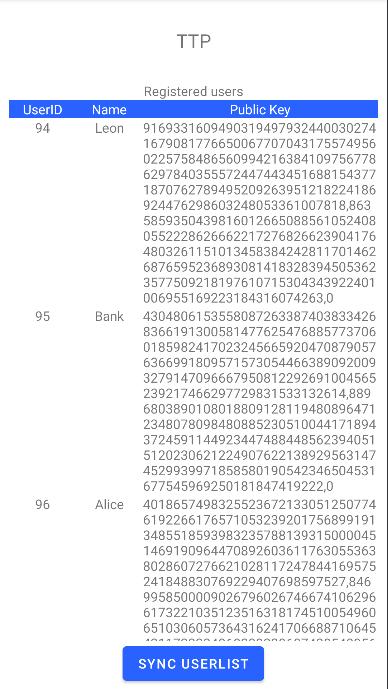}}
    \caption{Trusted Third Party} \label{fig:TTP}
  \end{subfigure}
  \hspace*{\fill} 
  \begin{subfigure}{0.31\textwidth}
    \fbox{\includegraphics[width=\linewidth, height=10cm]{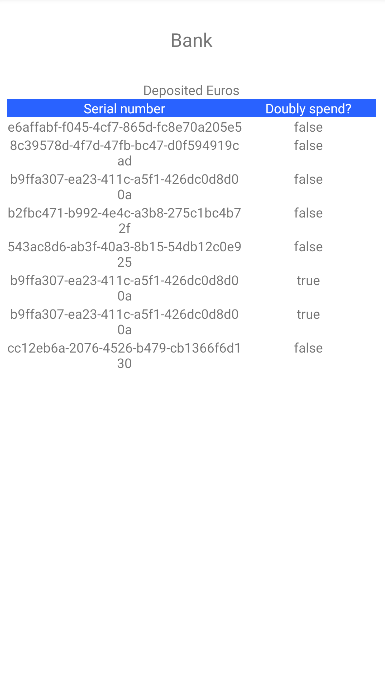}}
    \caption{Bank} \label{fig:bankScreen}
  \end{subfigure}%
  \hspace*{\fill}
  \begin{subfigure}{0.31\textwidth}
    \fbox{\includegraphics[width=\linewidth, height=10cm]{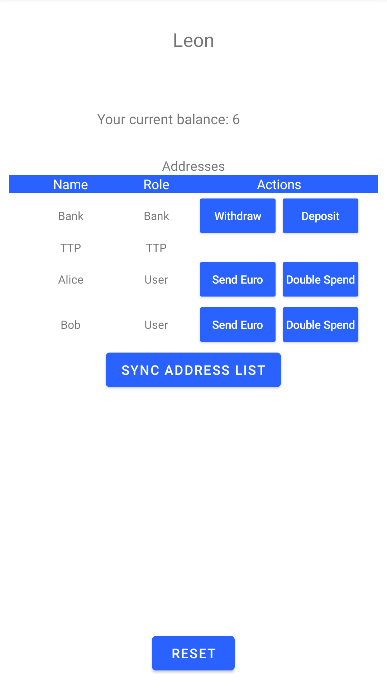}}
    \caption{User} \label{fig:userScreen}
  \end{subfigure}
  \caption{Screenshot of the application per role} \label{fig:screenshots}
\end{figure*}

The prototype is built as a mobile application, in which participants can select their role (TTP/Bank/User). Screenshots of the running application can be seen in Figure \ref{fig:screenshots}. For demonstration purposes, users (Figure \ref{fig:userScreen}) can easily double-spend digital euros. 

\fboxsep=0mm%padding thickness
\fboxrule=1pt%border thickness

The device communication is implemented using the Kotlin implementation of the peer-to-peer communication protocol IPv8\footnote{https://github.com/Tribler/kotlin-ipv8}. This library is also used during the transaction between users. Even though this library uses an internet connection, the (authenticated) messaging is done directly between two peers without a third party involved. As the library is solely used to transfer bytes between devices, the prototype can be extended with offline communication protocols, such as QR codes, Bluetooth or NFC.
As our fully offline constraint required transaction verification without needing a third party, the constraint is still satisfied and guaranteed this way. 

\subsection{Different curves using JPBC}
JPBC can be configured to use different types of underlying elliptic curves. 
This difference is the equation used to generate the bilinear map. 
Moreover, with JPBC it is also possible to set the security parameter giving more flexibility regarding the size of the group elements. 
The curves and their properties are listed in \cite{jpbcPaper}. 
The implementation has been tested with multiple underlying elliptic curves, both symmetric and asymmetric, and security parameters. 
For the different tested parameters, the protocol remained functional. 
This shows that the protocol is not tied to a specific curve type or security parameter. 

\subsection{Growth in size}
\label{GrowthExperiment}
As mentioned earlier, the size of the primary data structure to describe the digital euro must grow for each transaction. 
The growth size depends on the elliptic curve and security parameter used. 
To test the difference in growth, a test is done to measure the size of a serialized digital euro after each transaction, used in 50 transactions. 

\begin{table}[]
    \centering
    \begin{tabular}{|c|c|c|c|c|}
        \hline
        \textbf{Curve} & \textbf{$|G_1|$} &  \textbf{$|G_2|$} & \textbf{$|G_T|$} & \textbf{$|GS|$}\\
         \hline
         Type A & 128 & 128 & 128 & 1152 \\
         Type E & 256 & 256 & 256 & 2304 \\
         Type F & 40 & 80 & 240 & 720 \\
        \hline
    \end{tabular}
    \caption{Size of group elements and a single GS proof in bytes per curve type $(r=160)$.}
    \label{tab:groupElementSizes}
\end{table}

Three elliptic curves, A, E and F, were used for the test. To construct those curves, the curve generator of JPBC was used, with the default settings \footnote{http://gas.dia.unisa.it/projects/jpbc/docs/ecpg.html}.
Pairings of Type A and E are symmetric and the ones of Type F are asymmetric. Curves of Type A are used in Charm \cite{cryptoeprint:2011/617} and curves of Type F are BN-curves, introduced in \cite{barreto2005pairing}. BN-curves are used in Ethereum and Zcash \cite{aranha2023survey}.
As curves of Type E have larger elements, compared to Type A, and are less optimized, they are used less often \cite{kar2019systematization}. The sizes of the group elements \cite{kar2019systematization} can be found in Table \ref{tab:groupElementSizes}. The size of the GS proof is computed as $4 \cdot |G_1| + 4 \cdot |G_2| + |G_T|$.
During the test, the prime order used ($r$) remained constant ($r=160$).  

\begin{table}[]
    \centering
    \begin{tabular}{|c|c|c|c|}
        \hline
        \textbf{Curve} & \textbf{Initial size (kB)} &  \textbf{50 transactions (kB) } & \textbf{Growth (kB)} \\
         \hline
         Type A & 0.567 & 63.217 & 1.248 \\
         Type E & 0.823 & 114.673 & 2.272\\
         Type F & 0.391 & 41.441 & 0.816 \\
        \hline
    \end{tabular}
    \caption{Digital euro growth with transactions measured in serialized bytes $(r=160)$.}
    \label{tab:digitalEuroGrowth}
\end{table}

Table \ref{tab:digitalEuroGrowth} shows that the growth rate of the digital euro significantly depends on the elliptic curve. 
The asymmetric pairing (Type F) has the lowest growth rate, resulting from the smaller element sizes as found in Table \ref{tab:groupElementSizes}.
Due to the size of the elements, the digital euro will grow faster when elliptic curves of Type E are used, compared to curves of Type A. 

\begin{figure}
    \centering
    \includegraphics[width=\linewidth]{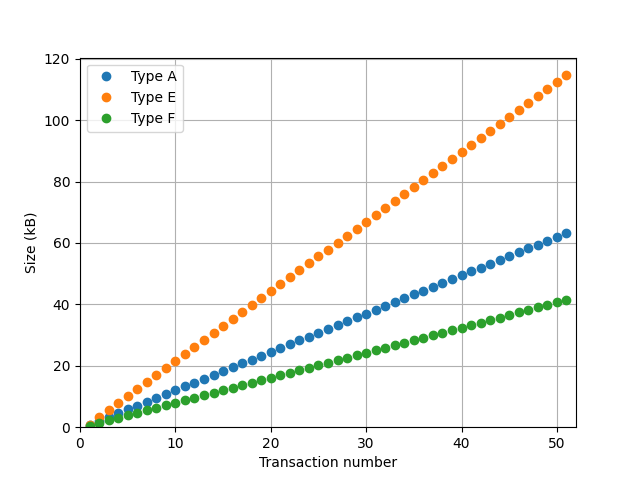}
    \caption{Serialized digital euro growth per transaction $(r=160)$.}
    \label{fig:eurogrotwh}
\end{figure}

The growth of the digital euro is constant for each transaction, hence the growth is linear to the number of transactions. 
In Figure \ref{fig:eurogrotwh}, the size of the digital euro is plotted after each transaction. 
The plot shows that the digital euro grows linearly with each transaction as expected from Equation \ref{sizeequation}.

Comparing the size of the expected growth ($|GS|$ from Table \ref{tab:groupElementSizes}) and the actual growth per transaction (Growth in Table \ref{tab:digitalEuroGrowth}), are similar in byte sizes. The difference between the expected and actual growth can result from serialization, overhead of the used list structure or how group elements are converted into bytes by JPBC. 
The first transaction, however, has a slightly larger growth ($+250$ bytes) due to the initialization of the list structure for the transaction proofs.

\subsection{Transaction verification performance}
The time it takes to verify a transaction is a major factor in adopting digital currencies and their usability in everyday transactions. 
For example, on average, a Bitcoin transaction takes 10 minutes to confirm and waiting for more confirmation blocks for larger transactions is recommended. 
As financial transactions are expected to be completed nearly instantly \cite{courtois2014could}, this payment option is unusable in most scenarios.

To test the transaction verification performance, a test measures the time it takes to verify a transaction using a digital euro for up to 50 transactions. To verify the final transaction the user must check three Schnorr signatures and the chain of 50 proofs. 

\begin{figure}
    \centering
    \includegraphics[width=1\linewidth]{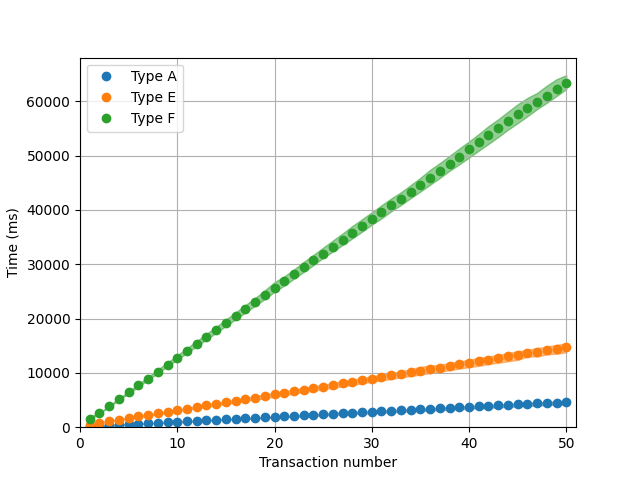}
    \caption{Transaction verification time $(r=160)$.}
    \label{fig:verificationPerformance}
\end{figure}

The test is set to measure the time it takes to verify a transaction. The digital euro is then used to create the next transaction details. This experiment is done using the same curves as mentioned in Section \ref{GrowthExperiment}. Furthermore, the experiment is executed 10 times for each curve. The results of this experiment are listed in Table \ref{tab:verificationperformance} and visualized in Figure \ref{fig:verificationPerformance}. In the figure, the minimum and maximum time needed to verify the transactions is also shown.

\begin{table*}[]
    \centering
    \begin{tabular}{|c|c|c|c|}
        \hline
        \textbf{Curve} & \textbf{First transaction (ms)} & \textbf{50th transaction} & \textbf{Average time increase per transaction (ms) } \\
         \hline
         Type A & 165 & 4545 & 91 \\
         Type E & 485 & 14470 & 292 \\
         Type F & 1530 & 62255 & 1259 \\
        \hline
    \end{tabular}
    \caption{Digital euro verification performance $(r=160)$.}
    \label{tab:verificationperformance}
\end{table*}

It is clear from the results that the verification of the proofs is the major part of the verification process. 
Similar to the growth of the digital euro, the time it takes to verify increases linearly with the number of transactions.
The elliptic curve used in the protocol significantly impacts the performance of the transaction verification. 

With the current implementation, the Type A curves outperform the other two curve types. 
Combining that with the results from the growth rate (Table \ref{tab:digitalEuroGrowth}), the curves of Type A seems to be the most favourable option.
Even though proofs created with the curves of Type F are more compact, the time it takes to compute the pairings makes them less usable after multiple transactions.

It is important to note that the implementation is not optimized for the best performance of the protocol. 
Therefore several steps can be taken to reduce the verification time of the transaction. 
The authors of JPBC mention that pairings in JPBC without preprocessing are roughly 5.5 times slower than the PBC \footnote{https://crypto.stanford.edu/pbc/} framework they ported to Java. Furthermore, the verification process is now single-threaded but could be parallelized as verifying the proofs themselves is not dependent on the other proofs. 
Other performance boosts could be preprocessing elements of the CRS or storing digital euros (partly) precomputed.
However, more research is needed for this.   

\section{Limitations and future work}
The current protocol relies on a TTP to revoke the anonymity of users in case double spending is detected. 
However, the TTP can revoke anyone's identity based on a single transaction. 
This makes it possible for a malicious TTP to fully trace transactions when it receives a digital euro with the full list of proofs. 
In most literature, the TTP requires two proofs of the double-spend transaction to revoke the user's anonymity. 
Even though this protocol offers more privacy and anonymity than the traditional banking system, a 'once concealed twice revealed' approach might be more desirable.
Such an approach might be feasible by using a different type of GS-proof. 
For example, by changing how the targets of the proofs are constructed. 
If it is possible to create the proofs such that two targets generated for the double-spending transaction would reveal the identity of the double-spender a commitment scheme that always hides the spender's identity can be used.

The ability to revoke the anonymity from one transaction has a legal advantage. 
When a perpetrator would only spend e-cash obtained through theft or a forced money transfer once, the perpetrator can be identified. 
Without this possibility, the perpetrator would not be identifiable from a valid transaction.

To further protect users' privacy, the CRS used in the protocol can be constructed by a collaboration of multiple parties. 
The ability of a single party to revoke the anonymity of all users is then removed. 
To revoke the anonymity of users all parties are needed. 

Another limitation is that users can recognize e-cash, which they had before. 
The signature and transaction proofs are not randomized with each transaction. 
Therefore if a user notices that it had the same e-cash before, it is possible to gain some knowledge regarding the traceability of the e-cash. 
This knowledge allows the user to link the receiver of the earlier transaction to the spender from whom the user received the e-cash and the number of transactions in between. 
This linkability could be avoided by randomizing both the signature and transaction proofs for every transfer as is done in \cite{baldimtsi2015anonymous} and \cite{bauer2021transferable}. 

More research is needed to determine which curve type is most optimal. 
This curve must balance the growth per transaction, the verification performance and the application's security. 
A more optimized version of the protocol is needed for this.
This optimization could be achieved by implementing the pairing in a more efficient framework. 
Moreover, other improvements could be preprocessing, parallelization and (partial) precomputation.
\section{Conclusion}
This thesis proposes an offline transferable e-cash scheme that could be used for the offline digital euro. Furthermore, this thesis introduces an open-source implementation of this scheme. The protocol offers fully offline transactions with transferable tokens and more privacy than current digital payment options or the earlier proposed EuroToken. 

Our implementation and tests show that the protocol is correct and that the number of transactions, in which the digital euro is used, affects the size of the digital euro and the transaction verification time in a linear manner. However, more research and optimizations are needed to find the best curve to use in the final implementation of the digital euro.

In conclusion, our implementation of the digital euro will enhance the digital payment ecosystem and solve the two main concerns with the current digital payment systems, namely the dependency on an online infrastructure and the fact that third parties can observe transactions. Our digital euro can make the economic system more durable and stable in areas with low coverage or during power outages, whilst providing more privacy than the current alternatives. 
\printbibliography

\end{document}